\documentclass{JHEP3}
\usepackage{graphicx}

\newcommand{\be}{\begin{equation}}
\newcommand{\ee}{\end{equation}}
\newcommand{\bea}{\begin{eqnarray}}
\newcommand{\eea}{\end{eqnarray}}

\def\IZ{\relax\ifmmode\hbox{Z\kern-.4em Z}\else{Z\kern-.4em Z}\fi}
\newcommand{\IS}{{\bf S}}

\newcommand{\non}{\nonumber \\}

\def\th{{\tilde h}}

\def\Sch{Schwarzschild}

\def\dl{\delta} 

\def\th{\theta}

\def\hlm{{\hat \lambda}}

\newcommand{\ltsim}{\protect\raisebox{-0.5ex}{$\:\stackrel{\textstyle <}
        {\sim}\:$}}
\newcommand{\gtsim}{\protect\raisebox{-0.5ex}{$\:\stackrel{\textstyle >}
        {\sim}\:$}}

\preprint{{\tt hep-th/0402216}}
\title{\center{}{A critical dimension in the black-string phase transition}}
\author{ Evgeny Sorkin
\\
 Racah Institute of Physics\\
 Hebrew University \\
 Jerusalem 91904,
 Israel\\
{\tt sorkin@phys.huji.ac.il}}

\abstract{
  In spacetimes with compact dimensions there exist several black
  object solutions including the black-hole and the black-string.
  These solutions may become unstable depending on their relative size
  and the relevant length scale set by the compact dimensions. The
  transition between these solutions raises puzzles and addresses
  fundamental questions such as topology change, uniquenesses and
  cosmic censorship.  Here, we consider black strings wrapped over the
  compact circle of a $d$-dimensional cylindrical spacetime.  We
  construct static perturbative non-uniform string solutions around
  the instability point of a uniform string.  First we compute the
  instability mass for a large range of dimensions, $d$, and find that
  it follows essentially an exponential law $\gamma^d$, where $\gamma$
  is a constant. Then we determine that there is a critical dimension,
  $d_*=13$, such that for $d\leq d_*$ the phase transition between the
  uniform and the non-uniform strings is of first order, while for
  $d>d_*$, it is, surprisingly, of higher order.}

\begin{document}
In 4d the static uncharged black hole(BH) solutions with a given mass
are stable and unique. However the  fundamental theory of nature, which as
now believed by many, is the string/M-theory contains more than four
dimensions. In this situation the phase space of  massive solutions
of General Relativity is much more rich and varied.  Several phases
of solutions exist and transitions between them may occur.  For
concreteness, we consider the background with a single compact
dimension, i.e. with the topology of a cylinder, ${\mathbb{R}}^{d-2,1}
\times \ {\bf S}^1$.  The coordinate along the compact direction is
denoted by $z$ and its asymptotic length is $L$. The problem is
characterized by a single dimensionless parameter
\be
\label{mu}
\mu:=G_d M / L^{d-3},
\ee
where $G_d$ is the $d$-dimensional gravitational constant and $M$ is
the mass.  

Gregory and Laflamme (GL) \cite{GL1,GL2} discovered that the uniform
black string (i.e. a $d-1$ \Sch \ solution times a circle, which is the
 large mass solution) develops a dynamical instability if the
compactification radius is ``too large''.  Their interpretation was
that the string decays to a single localized BH. In this case the
horizon pinches off and the central singularity becomes ``naked''.  By
now there is a rapidly growing amount of the literature on the subject
[3-21].
In particular the scenario of GL was questioned by Horowitz and Maeda
(HM) \cite{HM} who, on grounds of the classical ``no tear'' property
of the horizons, argued that horizon pinching is impossible and hence
a decaying string settles to another stable phase -- a non-uniform
black string(NUBS).  However, a (partial) evidence against that has
come from Gubser \cite{gubser} who in 5d studied perturbative NUBSs
emerging from the GL point. He showed that such solutions are too
massive and have too low an entropy to serve as an end-state of a
decaying critical string. Namely, the transition to this NUBS is of
first order and it is again unclear what state is accessed by the
classically decaying GL  string.  Wiseman \cite{Wiseman1} reached the same
conclusion by constructing the NUBS solutions numerically in 6d in a
fully non-linear regime\footnote{ In 5d, \cite{dynamic,deSmet} could
  be regarded as additional circumstantial evidences {\it contra} the
  HM claim.}.  However, in this paper we discover  that the  transition to 
NUBS  can be smooth depending on spacetime dimension.
 
Generalizing Gubser's 5d procedure \cite{gubser}, which is a version
of the ``marginal stability'' method, we construct numerically
$d$-dimensional static perturbative NUBS  solutions around the
 GL  point. First, we note that the GL instability  mass exhibits to
a good accuracy an exponential scaling with $d$.  Moreover, we find
that there is a critical dimension, $d_*=13$, below which the
uniform-nonuniform strings transition is of first order.  I.e. it is
qualitatively similar to what Gubser has found in 5d.  However, above
$d_*$ the NUBS solutions emerging from the instability point
have a lower mass and a larger entropy than those of the critical
string. Namely, the transition between the phases can be continuous\footnote{This is consistent with the prediction of a critical dimension
$\hat{d}=10$ at the ``merger point'' of this system, where the string
and the BH branches merge \cite{TopChange}.}.

Hence the NUBS state is accessible by an unstable uniform
string.  In this case, the horizon would not pinch off at the GL
point. Our result suggests, however, that the horizon
fragmentation during the classical decay can be avoided only for
$d>d_*$.  This is a rather curious development since the original HM
argument was dimension independent. It should be
noted, however, that the central issue of whether any unstable string
must decay to a string remains unresolved even for $d>d_*$.

 
 The most general ansatz for static black string solutions is
\bea 
\label{ansatz}
ds^2=-e^{2 A} f dt^2 &+&e^{2 B} \left( f^{-1} dr^2 +dz^2\right) +e^{2 C}
r^2 d\Omega_{d-3}^2, \non\
f&=&1-{1/r^{d-4}}, 
\eea
where $A,B$ and $C$ depend on $r,z$ only. When these functions vanish
the metric becomes that of a static uniform black string with the
horizon located at $r_h=1$.

Gubser \cite{gubser} has considered static NUBS solutions that differ
only perturbatively from a uniform black string.  Since the method was
described in detail in the original paper \cite{gubser} and then in
\cite{Wiseman1} we mention only the most important points. Gubser
developed a perturbation theory considering the expansion of the
metric functions in powers of $\hlm$. This $\hlm$ parametrizes the
NUBS branch that joins the GL point in the limit $\hlm
\rightarrow 0$. The expansion has the form\footnote{Gubser used the
  ``non-uniformity'' parameter, $\lambda:=1/2(R_{\rm max}/R_{\rm
    min}-1)$ where $R_{\rm max}$ and $R_{\rm min}$ refer to the
  z-dependent \Sch \ radius of the horizon.  Hence $\hlm$ coincides
  with this $\lambda$ only at the leading order.  It was shown
  subsequently in \cite{numericI,HO2} that a good order parameter that
  allows to put black strings and holes on the same phase diagram is
  not $\lambda$, which is undefined for the latter, but the scalar
  charge of the dilatonic field.  However, for our current purposes
  $\hlm$ may be left unspecified. }
\bea
\label{expansions}
 X &=& \sum_{n=0}^\infty \hlm^n X_n(r) \cos(n K z), \non
 X_n(r)&=& \sum_{p=0}^\infty \hlm^{2 p} X_{n,p}(r), ~~~
 K=\sum_{q=0}^\infty \hlm^{2 q} k_q,
\eea
for $X=A,B,C$  with $X_{0,0}=0$; and  $K = 2\pi /L$.

Upon substituting (\ref{expansions}) into the Einstein equations,
$R_{\mu\nu}=0$, a finite set of ODEs is generated at each order of the
expansion\footnote{See e.g.  \cite{numericII} for derivation of the
  Einstein equations in a very similar case.} .  Gubser's method is
very accurate up to the third order in $\hlm$.  Following the original
procedure we restrict our computations up to $O(\hlm^3)$.
Nevertheless, interesting results are already obtained here. Actually,
the third order is precisely what one needs to determine the
smoothness of a phase transition.

As discussed in \cite{gubser} the perturbation theory contains a
``scheme'' dependence that seems to correspond to different
parameterizations of the non-uniform branch.  Originally, fixing of
the ``scheme'' was achieved by fixing the constants
$c_{n,p}:=C_{n,p}(r_h)$. Still, other ``schemes'' can be used. For
example in \cite{Wiseman1} the asymptotic length of the compact circle
was held fixed, $K=const$, but the constants $c_{n,p}$ were allowed to
vary. In fact different ``schemes'' all produce the same
scheme-independent results, like e.g. the dimensionless mass
(\ref{mu}).  Here we choose to work in  the ``standard scheme'', as
it is referred in \cite{gubser}, by fixing $c_{n,p}=0$ for $n>1$ and
$c_{1,0}=1$.

Once the metric functions are known, various thermodynamical variables
can be computed. Asymptotically the spacetime (\ref{ansatz}) is
characterized by two charges\cite{numericI,HO2} -- the mass and the
tension of the black string.  By making a Kaluza-Klein reduction in
the $z$ direction, $X_{n,p}$ in (\ref{expansions}) are observed to be
massive modes for $n>0$ and they are massless otherwise. Only the
latter contribute to the asymptotic charges since the former decay
exponentially. Up to $O(\hlm^3)$ the relevant massless modes are
$X_{0,1}$.  Asymptotically, they fall off as inverse powers of $r$. We
denote the coefficients of the leading terms by $X_\infty$.  It is
convenient to define the {\it variation} of the charges of a
non-uniform string with respect to a uniform one.  According to
\cite{numericI} at the leading order these variations read\footnote{We
  use units in which $G_N:=G_d/L=1.$}
\bea 
\label{asympt_charges}
{\delta M/ M} &=&-2 \left[ A_\infty + B_\infty/(d-3)\
  \right] \hlm^2, \non {\delta {\cal T} / {\cal T}} &=&-2 \left[ A_\infty +(d-3) B_\infty
  \right]\hlm^2.
\eea
We also compute  the variation in the
temperature, $\delta T/T =\exp[A-B]-1$, and in the entropy,  $\delta S/S
=\exp[B-(d-3)C]-1 $, which are evaluated  at $r=1$.

Finally, defining the variation of $K$, $\delta K/K:=(k_1/k_0)
\hlm^2$, we determine the dimensionless, scheme-independent variables
by multiplying the dimensional quantities by  suitable powers of
$K$.  By doing so we obtain for our variables
\bea
\label{variables}
{\dl \mu/\mu} &=&{\dl M / M}+(d-4){\dl K / K} :=\eta_1
\hlm^2 +\dots,\non {\dl
  \tau/\tau}&=& {\delta {\cal
      T} / {\cal T}}+(d-4){\dl K / K}:=\tau_1
\hlm^2 +\dots, \non
{\dl\th / \th} &=& {\dl T/ T} -{\dl K / K}:=\theta_1
\hlm^2 +\dots,\non {\dl s
  / s} &=&{\dl S / S} +(d-3) {\dl K / K}:=s_1
\hlm^2 +\dots.
\eea

Incorporating the first law as in \cite{gubser} we evaluate the
entropy difference between the non-uniform and uniform strings with
the {\it same} mass
\bea
\label{entropy_dif}   
{S_{\rm non-uniform} \over S_{\rm uniform}} &=& 1 + \sigma_1 \hlm^2
+\sigma_2 \hlm^4 +\dots, \non
\sigma_1  =\eta_1-{d-4 \over d-3} {s_1} &,&
 \sigma_2 =-{d-3 \over 2(d-4)}\left(\th_1+{1\over d-4} {\eta_1}
 \right)\eta_1  .
\eea
The vanishing of $\sigma_1$ is ensured by the first law at the leading
order (where $L=const$)\cite{gubser}. We verified that to a good
($\ltsim 1\%$) accuracy, $\sigma_1\approx 0$ for our solutions. Thus,
the entropy difference (\ref{entropy_dif}) arises only at $O(\hlm^4)$.

At each order of $\hlm$ we solved the ODEs
numerically\cite{WisemansMathematica}. We were able to exactly
reproduce the numbers found so far in the literature: for 5d in
\cite{gubser} and for 6d in \cite{Wiseman1}. An indication of the
accuracy of the method is gained by varying the ``scheme''
\cite{gubser},  by altering $c_{0,1}=0,\pm 1$. The resulting
variation in (\ref{variables},\ref{entropy_dif}) gives an idea of the
numerical uncertainty.  For small $d$'s the accuracy of our
calculation is high, being about $0.5 \%$ in $\eta_1$ and $1\%$ in
$\sigma_2$.  For larger $d$'s the method is somewhat less accurate,
yielding $5\%$ and $6\%$ variations in $\eta_1$ and $\sigma_2$
respectively, for $d=16$. This has to do with the steep asymptotic
fall off of  $A$ and $B$ in which the leading terms decay
as $r^{-(d-4)}$, while $C$ falls off only as $1/r$ (in 5d the fall off
is $\log(r)/r$).  Hence, the
accuracy in extracting the coefficients $A_\infty, B_\infty$, that
contribute to $\eta_1$ and $\sigma_2$, decreases for large $d$.
 
\begin{figure}[t!]
\centering
\noindent
\includegraphics[width=10cm]{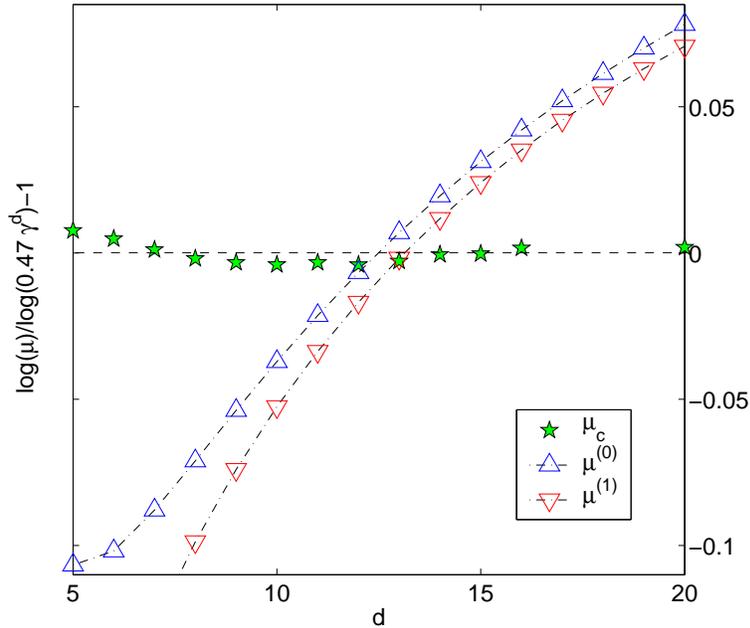}
\caption[]{The relative {\it difference} between the mass  and the fit
  (\ref{mu_crit}), $0.47 \gamma ^d$, as a function of $d$.  For
  $\mu_c$ this difference is zero with the spread of about $0.8\%$
  magnitude, giving approximately $2.1\%$ variations in $\mu_c$
  itself.}
\label{fig_mu}
\end{figure}

{\it The critical  mass.}
The calculation in the linear order in $\hlm$ yields the mass of the
critical string, since the leading order of (\ref{expansions})
corresponds to the static GL mode. We performed the calculations in
$d=5,\dots,16,20,30$ and $50$.  For $d\leq 10$ we confirm a very good
agreement with the original GL results\cite{GL1}, presented in their
FIG. 1. Note, however, that the methods are very different. For the
entire range of $d$ we find that the critical mass is remarkably well
approximated by
\be
\label{mu_crit}
\mu_c \propto \gamma^d
\ee
with $\gamma \simeq 0.686$ being a constant, and the prefactor is
approximately $0.47$, for the specific definition of  mass
(\ref{mu}). In FIG. \ref{fig_mu} we plot the relative {\it
  difference} between the logarithm of the critical mass and the fit
(\ref{mu_crit}). It is clearly seen that $\log(\mu_c)$ is linear for
all $d$.  There is still room for a weak $d$-dependence, of order
$2.1\%$, around the dominant scaling (\ref{mu_crit}).  We, however,
could not extract this residual dependence.

To get an insight into this behavior (\ref{mu_crit}) we compute the
mass of a uniform black string whose entropy is equal to that of a
single BH with the same mass.  First we compare the entropy of
the black string with that of a d-dimensional \Sch \ BH.  The
corresponding entropies read $S_{\rm BH}^{(0)}=A_{d-2}/(4G_d)$ and $
S_{\rm BStr}=A_{d-3} L /(4G_d)$ where $A_d:= \Omega_d [16\pi M /(d \,
\Omega_d)]^{d/(d-1)}$ and 
 $\Omega_{d}$ is the surface area of a unit $\IS^{d}$ sphere.
Equating these, $S_{\rm BH}^{(0)}(\mu)=S_{\rm BStr}(\mu)$, we solve for the mass
\be 
\label{mu_0}
\mu^{(0)}= {1 \over 16 \pi}{\Omega_{d-3}^{d-3} \over \Omega_{d-2}^{d-4} }
  {(d-3)^{(d-3)( d-3)} \over  (d-2)^{(d-2 )(d-4)}}.
\ee

Actually, we can do slightly better by using the analytical formula
for the entropy of  {\it small} BHs on cylinders derived
recently in \cite{Harmark}
\be 
\label{entropy_Harmark}
S_{\rm BH}^{(1)}=S_{\rm BH}^{(0)}\left[1+{\zeta(d-3) 16 \pi  \mu \over 2(d-3)
    \Omega_{d-2}} +O(\mu^2)\right].
\ee
where $\zeta(x)$ is Riemann's zeta-function. This formula reflects the
leading order corrections to the \Sch \ metric due to
compactification\footnote{The perturbation theory is constructed in
 powers  of $\mu \ll 1$.}. It implies that for a given mass the
entropy of a ``caged black hole''(a BH in a compactified
spacetime) is larger than the entropy of a \Sch \ BH. The mass
$\mu^{(1)}$ corresponding to equality of the entropies is then
obtained by solving the equation $S_{\rm BH}^{(1)}(\mu)=S_{\rm
  BStr}(\mu)$.

We add to  FIG. \ref{fig_mu} the plots for these masses. In contrast
to $\log(\mu_c)$ the logarithms of $\mu^{(0)}$ and $\mu^{(1)}$ have a
non-linear dependence on the dimension for small $d$'s. They do,
however, become linear (with a different slope) for $d\gg 10$.  Here we
already see a hint of a critical dimension -- looking at the
difference between $\mu_c$ and its estimator (either $\mu^{(0)}$ or
$\mu^{(1)}$) one notices a change of sign at about $d\sim 12.5$.  This
suggest that for $d\gtsim 13$ the BH state is entropically
favorable over the string state only  for $\mu<\mu_c$.

\begin{figure}[t!]
\centering
\noindent
\includegraphics[width=10cm]{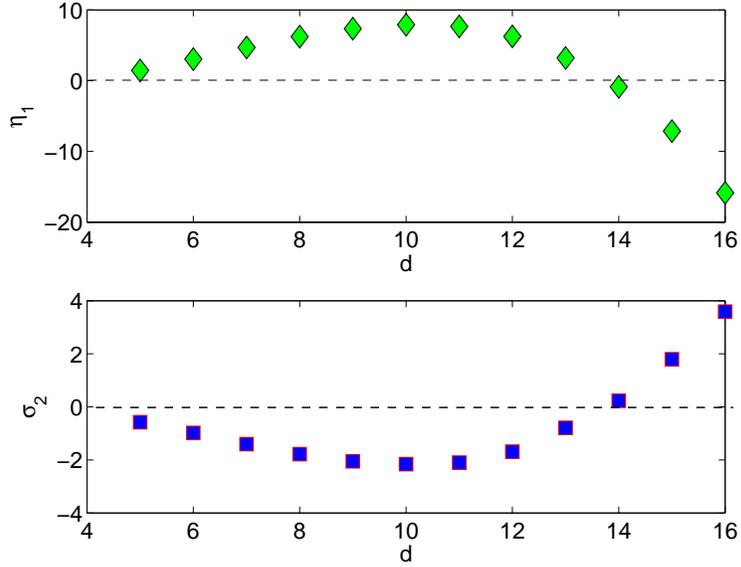}
\caption[]{The trends  in the mass, $\mu_{\rm non-uniform} /\mu_{\rm
    uniform} :=1+\eta_1 \hlm^2 + \dots$, and the entropy, $S_{\rm
    non-uniform} /S_{\rm uniform} :=1+\sigma_2 \hlm^4 + \dots$,
  shifts between uniform and non-uniform black strings.  The key
  result is the sign change of $\eta_1$ and $ \sigma_2$ above $d_* =
  13$. }
\label{fig_variables}
\end{figure}

{\it From a sudden to a smooth phase transition.}  Performing the
computation in higher orders, up to $O(\hlm^3)$, we obtain the
variation in the variables (\ref{variables}) and entropy
(\ref{entropy_dif}). The results for $\eta_1$ and $\sigma_2$ are
depicted in FIG. \ref{fig_variables}.
One observes that $\eta_1$ is initially positive for $d=5$, reaches a
maximum at $d=10$, and becomes negative for $d>13$. Then it continues
to decrease and in fact it drops increasingly faster with $d$, as
indicated by the growing distances between subsequent points in the
graph.  The pattern for $\sigma_2$ is similar but with the opposite
sign\footnote{ In fact we also did the computation in $d=20$ finding
  the same trends. However, the numerical errors were of order $20\%$
  so we regard this case as an indicative only.
}.

The key phenomena is the appearance of a critical dimension, $d_*=13$,
above which the perturbative non-uniform strings  are less
massive than the marginal GL string. Moreover, their entropy is larger
than the entropy of the uniform string with the same mass. It is 
important that $\eta_1$ and $\sigma_2$ change signs
simultaneously.

As for the other variables, we find that the trend in the entropy
shift, $s_1$, is qualitatively similar to the behavior of $\eta_1$ --
it is positive for $d\leq d_*$ and it becomes negative above $d_*$.
For the variation of the temperature we note that below $d_*$ the
NUBS is ``cooler'' than the uniform one and above $d_*$
it is ``hotter''. We find that the tension of the non-uniform strings
is lower than that of uniform ones.  This is in tune
with the expectation that the uniform black string has a maximal
tension, and that the tension vanishes for  small black
holes \cite{numericI,Myers}.  In addition, we observe the ratios
$\eta_1 /\tau_1 $ and $\eta_1 /s_1 $ to be discontinuous near $d_*$.
Note also that in FIG. \ref{fig_variables} we plot the coefficients
of the mass and the entropy shifts. To obtain the physical variations
these and others coefficients must be multiplied by suitable powers of
$\hlm$.

  
To summarize. While we have found the dependence of the critical mass
on the dimension we do not have at present an explanation for the
scaling (\ref{mu_crit}). We believe, it gives us some insight
into the nature of the GL instability and it probably is connected with the
thermodynamical instability of the system\cite{thermo}.  However, it
is the appearance of a critical dimension, $d_*$, that can perhaps be
regarded as our main result. It implies that above $d_*$ the critical
string can smoothly evolve into the NUBS phase.  For $d\leq d_*$ the
transition between the two phases is of first order.

The continuous transition above $d_*$ suggests that the NUBS phase can
be a natural end state of the GL instability.  Indeed, a uniform
string loosing its mass by evaporation and encountering the
instability at $\mu_c$ can smoothly evolve to the non-uniform state
keeping its singularity covered by the horizon.  Already from FIG.
\ref{fig_mu} it could be inferred that above $d\gtsim 13$ there can be a
branch of solutions between the uniform strings and the BHs. We
believe that the NUBS state is a reasonable candidate for this
``missing link''.

As the mass is further radiated away two scenarios may be proposed:
(1) The NUBS branch extends to an arbitrary small mass. A black string
evolves along this branch probably increasing its non-uniformity all
the way down to zero mass.  In this case the cosmic censorship would
be held(at least until the final stages of evaporation); (2) A NUBS
becomes unstable at a finite mass where the horizon fragments and a
localized BH forms. This may lead to a compromize of the cosmic
censorship, much like in the $d\leq d_*$ case but for a mass smaller
than $\mu_c$.  The transition between a NUBS and a BH can be sudden or
smooth depending on the relative values of the instability masses for
these states. Note that a NUBS branch that extends to zero mass or
becomes unstable even earlier on a phase diagram is conceptually the
same. The main difference is whether the naked singularity shows up
before the end of evaporation or not.

To address these intriguing issues it would be a very interesting
future task to construct in a fully non-linear regime, like in
\cite{Wiseman1}, the branch of NUBSs that we found here. In
particular, it is interesting to determine for how low a mass this
branch drops, would the horizon try to pinch off forming a cone-like
``waist'' \cite{TopChange,KolWiseman} and whether the topology tends
to change. In addition we expect that a time evolution of the critical
string, like in \cite{dynamic}, should confirm a nice decay for
$d>d_*$.

In this work we have considered black strings in a cylindrical
spacetime.  We believe that the critical dimension phenomena is
general and will hold for more general backgrounds with additional
compact dimensions even if the specific value $d_*=13$ would change.

I thank B. Kol and T.  Piran for  stimulating discussions and 
valuable remarks on the manuscript and Mu-In Park for pointing out 
an error in the $\mu^{(1)}$ calculation.


\end{document}